\documentclass[aps,pre,twocolumn,groupedaddress,showpacs,showkeys,floatfix]{revtex4-1}
\usepackage{wrapfig}
\usepackage{graphicx}
\usepackage{colordvi}
\usepackage{color}
\usepackage{amsmath}
\usepackage{amssymb}

\newcommand{\hq}{h}
\newcommand{\wf}{w}
\newcommand{\Ww}{W}
\newcommand{\Hh}{H}

\begin{document}
\title{Multicanonical sampling of rare events in random matrices}
 
\author{Nen Saito}
\affiliation{
Graduate School of Science and Cybermedia Center, Osaka University,
Toyonaka, Osaka 560-0043, Japan}
\email[]{saito@cp.cmc.osaka-u.ac.jp}
\author{Yukito Iba}
\affiliation{The Institute of Statistical Mathematics,
10-3 Midorimachi, Tachikawa, Tokyo 190-8562, Japan}
\email[]{iba@ism.ac.jp} 
\author{Koji Hukushima}
\affiliation{
Department of Basic Science, University of Tokyo,
3-8-1 Komaba, Meguro-ku, Tokyo 153-8902, Japan}
\email[]{hukusima@phys.c.u-tokyo.ac.jp}

\date{\today}
\begin{abstract}
A  method
based on multicanonical Monte Carlo 
is applied to the calculation of 
large deviations in the largest eigenvalue of random matrices.
The method is successfully tested 
with the Gaussian orthogonal ensemble (GOE), 
sparse random matrices, and matrices whose components are 
subject to uniform density. Specifically,
the probability that all
eigenvalues of a matrix are negative is
estimated in these cases down to the values 
of $\sim 10^{-200}$,
a region where simple
random sampling is ineffective.
The method can be applied to any ensemble
of matrices and used for sampling rare
events characterized by any statistics.
\end{abstract}
\maketitle

\section{Introduction}

Rare events caused by rare realization of impurities
often govern the properties of random 
systems and play an essential role
in their study. Numerical computation
of the probabilities of rare events
is, however, computationally expensive.
When the probability
takes very small values, say, $10^{-15}$ or less, it is
virtually impossible to calculate the correct
probability value by simple random sampling.

Recently, approaches based 
on dynamic Monte Carlo (Markov chain
Monte Carlo)~\cite{metropolis1953equation, 
landau2005guide, *newman1999monte, *xgilks1996markov} 
have been shown to be useful 
for sampling rare events 
and calculating large deviations
in the corresponding statistics. 
The novelty of the approach is that 
a dynamic Monte Carlo algorithm is used for
calculating sample averages over 
configurations of impurities, instead of
computing thermal averages.
Successful examples in physics include
applications in spin glass~\cite{koerner2006probing, matsuda2008distribution}, diluted 
magnets~\cite{hukushima2008monte}, and  directed random walks in random 
media~\cite{monthus2006probing}. Some references~\cite{hartmann2002sampling, 
holzloohner2003use, *holzlohner2005evaluation, yukito2008testing, driscoll2007searching} 
also discuss applications in information processing and 
other engineering problems. 

The aim of this paper is to apply the method to
sample rare events in random matrices.
Random matrices have been a classical 
subject with a number of applications
 in physics and other fields~\cite{wigner1955characteristic, *wigner1958distribution, dyson1996selected, mehta2004random}.
Specifically,  large deviations
in the maximum eigenvalue of random matrices
is a subject of recent interest in various
fields such as ecology \cite{may}, 
cosmology~\cite{aazami2006cosmology}, mathematical
statistics~\cite{roy1957some}, and information compression~\cite{candes2006near}. 
The tail of the distribution of the maximum eigenvalue
is important because
it gives the probability of all eigenvalues
being negative, which is often related to the stability
condition of complicated systems~\cite{aazami2006cosmology, may}.

A  well-known study by Tracy and Widom
established the celebrated ``1/6 law"~\cite{tracy1994level,*tracy1996orthogonal}
on small deviations 
in the maximum eigenvalue of random matrices.
On the other hand, analytical studies~\cite{dean2006large, dean2008extreme, 
majumdar2009large, *vivo2007large} of  large deviations
give estimations of the tails of probabilities in special  cases such as  the
Gaussian orthogonal ensemble (GOE) and 
ensemble of random Wishart matrices.  However,
the techniques based on  the Coulomb gas representation
are difficult to generalize to ensembles
with other distributions of the components.
Other results by mathematicians and physicists
are also limited to special ensembles and/or give only
the upper bound of the probabilities~\cite{candes2006near}.
Thus, an efficient numerical approach that
enables exploration of  extreme tails of density
is necessary.

We propose a method based on multicanonical 
Monte Carlo~\cite{berg1991multicanonical, 
*berg1992multicanonical,  berg1992new}
as  a promising approach to the problem.
As we will show in this study,
quantitative results  are obtained in examples
of sparse random matrices and matrices
whose components are subject to 
the uniform density.
A similar method is used 
in~\cite{driscoll2007searching} 
to calculate large deviations in the \textit{growth ratio} of 
matrices. 
The paper~\cite{driscoll2007searching}, however, focuses on
applications in numerical analysis and 
does not compute large deviations in the largest eigenvalues.

The organization of this paper is as follows:  In Sec.~\ref{sec:multi}, we
summarize the multicanonical Monte Carlo algorithm. In Sec.~\ref{sec:dev},
we discuss how multicanonical Monte Carlo is 
used to calculate large deviations. In Sec.~\ref{sec:dev2}, the
results of numerical experiments on the tails of the
distribution of the largest eigenvalues are shown.
 Sec.~\ref{sec:prob} covers the computation of 
the probability that all eigenvalues are negative; as noted above, 
this is a typical application of the proposed 
method. In. Sec.~\ref{sec:sparse}, sparse matrices are treated. 
In Sec.~\ref{sec:con}, concluding remarks are given.

\section{Multicanonical Monte Carlo} \label{sec:multi}

Let us summarize the idea of multicanonical 
Monte Carlo~\cite{berg1991multicanonical, 
*berg1992multicanonical, berg1992new}.
Assuming the energy $E(x)$ of a state $x$, our task 
is to calculate the density $D(E)$ of states
defined by
\begin{equation}\label{ddd}
D(E)= \int \delta(E(x)-E) \, dx,
\end{equation}
where $\delta$ is the Dirac $\delta$-function, and 
$\int \cdots dx$ denotes a multiple integral in the
space of states $x$.

A key quantity of multicanonical Monte Carlo 
is the weight function $\wf(E)$ of the energy $E$.
Performing dynamic Monte Carlo sampling with the weight
$\wf(E(x))$, we modify $\wf(E)$ step-by-step
until the marginal density $\hq(E)$
of $E$ is almost flat in 
a prescribed interval $E_{\min}<E<E_{\max}$.
The initial form of $\wf(E)$ is arbitrary, and we can start,
for example, from a constant function.
Several methods are proposed for optimizing
a univariate function $\wf(E)$, among which a method
proposed by Wang and Landau~\cite{wang2001efficient, *wang2001determining}
is most useful and 
used in this paper (see the appendix). 
After a weight function 
$\wf^*(E)$ that gives a sufficiently flat $\hq(E)$ 
is obtained, we compute an accurate 
estimate $\hq^*(E)$ of $\hq(E)$ 
by a long simulation run
with the weight $\wf^*(E(x))$. Then, $D(E)$
is estimated by the relation
$$
D(E) \propto \frac{\hq^*(E)}{\wf^*(E)} {}_, \qquad (E_{\min}<E<E_{\max})
{}_.
$$

This simple algorithm has significant advantages
over conventional methods for estimation 
of $D(E)$. First, it realizes
accurate sampling of tails of 
$D(E)$ without estimating densities in the
high-dimensional state space of $x$.
Second, when we include the region of $E$ with large
values of $D(E)$ in the interval $E_{\min}<E<E_{\max}$, 
the mixing of dynamic Monte Carlo is dramatically facilitated.
This ``annealing'' effect is the reason that
multicanonical Monte Carlo is successfully used
to calculate thermal averages at low temperatures
in the studies of spin glass~\cite{berg1992new} and 
biomolecules~\cite{hansmann1993prediction}.

\section{Large deviations in the largest eigenvalues} \label{sec:dev}

An essential observation in the present approach is 
that the energy $E$ in multicanonical Monte Carlo 
need not be an energy in the ordinary sense.
That is, we can substitute for $E$ any quantity for which
we are interested in its rare fluctuations 
or large deviations from the average; 
similar approaches to other problems are found in~\cite{
holzloohner2003use, *holzlohner2005evaluation, driscoll2007searching,
hukushima2008monte, yukito2008testing, matsuda2008distribution}.

In this study, we regard the maximum eigenvalue $\lambda_1(x)$
of a matrix $x$ as a fictitious ``energy'' of the state $x$.
Also, we can introduce an underlying 
density $p(x)$ that gives the 
probability of $x$ under random sampling.
While $p(x)$ is the uniform density  in statistical mechanics,
$p(x)$  in the present case 
characterizes an ensemble of  matrices.  
Hereafter, we denote the size and the 
$(i,j)$-component of the matrix $x$ as $N$ and $x_{ij}$, respectively.
Also, we assume  the factorization $p(x)= \prod_{ij} p_{ij}(x_{ij})$; 
when we consider an ensemble of symmetric matrices,
$x_{ji}=x_{ij}$ and the product is taken for $i \leq j$.
The normalized density  $D(\lambda_1)$ of the states is written as
\begin{equation}\label{ddd1}
D(\lambda_1)= \int \delta(\lambda_1(x)-\lambda_1)
p(x) \, dx,
\end{equation}
where we replace $E(x)$ in~(\ref{ddd}) 
with $\lambda_1(x)$. $D(\lambda_1)$ is simply
the probability distribution of $\lambda_1$, whose
extreme tails we are interested in.

Now the application of multicanonical Monte Carlo
is straightforward. We employ a Metropolis-Hastings
algorithm~\cite{hastings1970monte} to generate samples according to 
the weight $\wf(\lambda_1(x))p(x)$.
A single component $x_{ij}$ of the random matrix $x$ 
is chosen and changed at each step; in ensembles of symmetric matrices,
$x_{ji}$ should also be changed if $i \neq j$, 
which is necessary to keep the symmetry of the matrix. 
The candidate $x_{ij}^{new}$ of $x_{ij}$ is generated
according to  the proposal 
density $r_{ij}(x_{ij}^{new}|x^{old})$, where $x^{old}$
is the current value of $x$; 
$x_{ij}^{new}$  is accepted  if and only if the Metropolis ratio 
$$
R= \frac{p_{ij}(x_{ij}^{new})}{p_{ij}(x_{ij}^{old})} 
\frac{r_{ij}(x_{ij}^{old}|x^{new})}{r_{ij}(x_{ij}^{new}|x^{old})}
\frac{\wf(\lambda_1(x^{new}))}{\wf(\lambda_1(x^{old}))}
$$
is smaller than a random number uniformly distributed in $(0,1]$.
Repeating this procedure, the function $\wf(\lambda_1)$ 
is tuned by the method of 
Wang-Landau~\cite{wang2001efficient, *wang2001determining}, whose 
details are given in the appendix.
Once a weight 
function $\wf^*(\lambda_1)$ that gives a sufficiently flat $\hq(\lambda_1)$ 
is obtained,
we estimate $D(\lambda_1)$ using
the formula
$$
D(\lambda_1) 
\propto \frac{\hq^*(\lambda_1)}
{\wf^*(\lambda_1)} {}_, 
\qquad (\lambda_1^{\min}<\lambda_1<\lambda_1^{\max})
{}_,
$$
where $\hq^*(\lambda_1)$ is the density
of $\lambda_1$ estimated by
a long run with the 
fixed weight function $\wf^*(\lambda_1)$.

A simple choice  of  the proposal density 
is $r_{ij}(x_{ij}^{new}|x^{old})=p_{ij}(x_{ij}^{new})$, 
which results in  a simple form of the Metropolis ratio
$$
R=\frac{\wf(\lambda_1(x^{new}))}{\wf(\lambda_1(x^{old}))}.
$$
However, this choice may not be adequate in some cases
where the support of the densities $p_{ij}(x_{ij})$ is not finite,
because very large deviations in an element $x_{ij}$
can be relevant for large deviations in the largest eigenvalue.
In these cases, if 
candidates with required values of $x_{ij}$
are rarely generated by the proposal
density $p_{ij}(x_{ij}^{new})$, the algorithm fails.
Typical cases arise when we examine extreme lower
tails of the distribution for relatively small matrix size $N$.

An alternative choice is to use 
$r_{ij}(x_{ij}^{new}|x^{old})=\tilde{r}_{ij}(x_{ij}^{new}-x_{ij}^{old})$,
where $\tilde{r}_{ij}( \cdot)$ is an even function; hereafter  we will call
an algorithm using this proposal density 
as a {\it random walk scheme}. 
The Metropolis ratio is given by
$$
R=\frac{p_{ij}(x_{ij}^{new})}
{p_{ij}(x_{ij}^{old})} \frac{\wf(\lambda_1(x^{new}))}{\wf(\lambda_1(x^{old}))}.
$$
With this choice, we can avoid 
the above-mentioned difficulty, because candidates
with any  large $|x_{ij}|$ can be generated 
in a step-by-step way, if they are accepted in intermediate
steps.

Throughout this study,  we have tested both choice 
of the proposal densities, but the only example in this paper
where the proposal density $r_{ij}(x_{ij}^{new}|x^{old})
=p_{ij}(x_{ij}^{new})$ gives inadequate results is the one
shown in Fig.~\ref{fig:new}, where we should calculate extremely
small probability for $N=10$ and $20$. In all other cases 
we have tested, no significant differences are found.

\section{\protect Computation of Density $D(\lambda_1)$} \label{sec:dev2}

We test the proposed method with 
the Gaussian orthogonal ensemble (GOE); 
GOE is an ensemble of real symmetric matrices 
whose entries are independent Gaussian variables~\cite{mehta2004random}.
In the following experiments, the variances of the diagonal and off-diagonal components
are $1$ and $1/2$, respectively, while means are all zero.
The Householder method is used to diagonalize the matrix in each step;
it is also used in other examples in this paper. 
We employ two different 
forms of the proposal density:
(1)~$r_{ij}(x_{ij}^{new}|x^{old})=p_{ij}(x_{ij}^{new})$ 
and  (2)~$r_{ij}(x_{ij}^{new}|x^{old})=p_{ij}(x_{ij}^{new}-x_{ij}^{old})$; the latter is
a special case of the random walk scheme, where $\tilde{r}_{ij}(\cdot)
=p_{ij}(\cdot)$~\footnote{This choice of the proposal density
is somewhat arbitrary;  for example, we
can use  Gaussian densities with different variances
and zero mean}.  

\begin{figure}
\begin{center}
\includegraphics{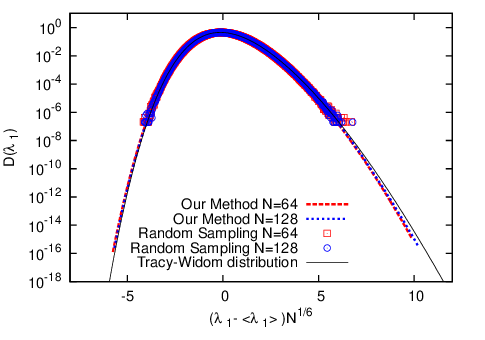}

\small
  \caption{Density $D(\lambda_1)$ in  GOE.
Results of the proposed method and 
a simple random sampling method are compared for
$N=64$ and $128$. Both almost overlap
with $N^{1/6}$ scaling. The symbols {\tiny $\Box$} and $\circ$ appear only in
 the region where 
simple random sampling gives meaningful results.
The Tracy-Widom distribution
is shown by the solid curve~\footnote{Figures in the tables at
{\tt http://www-m5.ma.tum.de/KPZ/} are used to draw curves
in Figs.~\ref{fig:1} and \ref{fig:new}}.
}
  \label{fig:1}
\end{center}  
\end{figure}

In Fig.~\ref{fig:1}, results estimated with 
the proposed method  with $r_{ij}(x_{ij}^{new}|x^{old})=p_{ij}(x_{ij}^{new})$
are compared with 
the corresponding results of simple random sampling.
The total  number of matrix diagonalizations  is $4.5\times10^8$
for $N=64$ and $2.25\times10^8$ for $N=128$;
they are the same in both of the proposed method
and simple random sampling. 
In  the proposed method, two third of them
are used  to optimize the weight, while  the rest
is used  to calculate the estimates.
We confirmed that modification factors
in the Wang-Landau method are sufficiently 
close to unity at the end of the weight optimization.
For $N=64$, we also apply the random walk scheme 
$r_{ij}(x_{ij}^{new}|x^{old})=p_{ij}(x_{ij}^{new}-x_{ij}^{old})$
and obtain the same result, but the computational time increases.
The results in Fig.~\ref{fig:1} show that the proposed method 
enables us to estimate the tails of the
density down to $\sim 10^{-15}$, 
which is scarcely sampled by the simple random sampling.

\begin{figure}
\begin{center}
\includegraphics{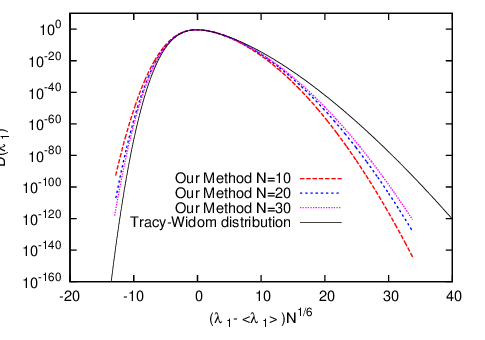}

\small
  \caption{
Density $D(\lambda_1)$ in  GOE for smaller $N$'s.
Results of the proposed method are shown
for $N=10, 20$ and $30$. The Tracy-Widom distribution
is shown by the solid curve.
}
  \label{fig:new}
\end{center}  
\end{figure}

For smaller $N$,  diagonalization of matrices takes less time and
even much smaller probabilities are computed. 
Figure~\ref{fig:new} shows results for $N=10, 20$ and $30$, where 
probabilities are computed down to $\sim 10^{-100}$;
deviations from the Tracy-Widom distribution
in the tails of the distributions become evident.
The random walk scheme is applied;
as we already mentioned in Sec.~\ref{sec:dev}, it is the only
case in this paper that the choice $r_{ij}(x_{ij}^{new}|x^{old})=p_{ij}(x_{ij}^{new})$ 
does not work.
The total number of matrix diagonalizations is $3\times10^9$
for $N=10$ and $4.5\times10^9$ for $N=20$ and $30$;
two third of them are used  to optimize the weight.

\section{The probability that all eigenvalues are negative} \label{sec:prob}

The proposed strategy also allows 
us to calculate the probability  $P(\forall i, \lambda_i < 0)$
that all eigenvalues of a random matrix are negative, which 
is important in applications 
in a variety of fields~\cite{aazami2006cosmology, may}.
Using the relation $\forall i, \lambda_i \leq \lambda_1$,
this probability is calculated by
$$
P(\forall i, \lambda_i < 0)=
{\int_{\lambda_1^{\min}}^{0}
D(\lambda_1) \,d\lambda_1} {}_.
$$
Here we assume that the density $D(\lambda_1)$ of the maximum
eigenvalue is estimated 
in an interval $[\lambda_1^{\min},\lambda_1^{\max}]$ by the proposed
method. 
The probabilities  {$P(\lambda_1<\lambda_1^{\min})$}
and {$P(\lambda_1^{\max}<\lambda_1)$} are
also assumed to be negligibly smaller than {$P(\lambda_1^{\min}<\lambda_1<0)$} and
{$P(\lambda_1^{\min}<\lambda_1<\lambda_1^{\max})$}, respectively.

The probability that all eigenvalues are positive can also
calculated with a similar way; it coincides with the probability
that all eigenvalues are negative when the distribution of
components is symmetric with respect to the origin.

First, we test the proposed method with GOE, where
the asymptotic behavior of $P(\forall i, \lambda_i < 0)$ for large $N$
is given by Dean and Majumdar \cite{dean2006large, dean2008extreme} as
$$
P(\forall i, \lambda_i < 0)\sim \exp\left ( -aN^2  \right ) {}_,
$$
where $a=\frac{\ln 3}{4} =0.274653\cdots$ .
This expression is derived
by interpreting the eigenvalues as a Coulomb gas,
a method that obviously does not apply general
distribution of components of matrices.

Confirming the result by numerical methods
is difficult because we should sample very rare events
to estimate the tails of the distribution.
Dean and Majumdar~\cite{dean2006large, dean2008extreme} 
(and Aazami and Easther~\cite{aazami2006cosmology})
provided numerical results
by simple random sampling, but their results are limited 
to small $N$, such as
$N=7$ in~\cite{dean2006large, dean2008extreme} 
 ($N=8$ with an additional assumption~\cite{dean2008extreme}). 
Dean and Majumdar also did numerical computation
up to $N=35$  based on the Coulomb gas representation; 
their computation does not, however, provide an independent check
to the theory and cannot be generalized to an arbitrary ensemble.

\begin{figure}
\begin{center}
\includegraphics{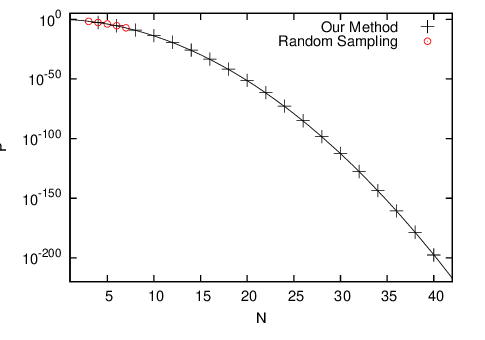}
\small
  \caption{Probability $P(\forall i, \lambda_i < 0)$  for GOE
versus size $N$ of the matrices.
The results of the proposed method (+) and the simple
random sampling method ($\odot$) are shown.
The results of simple random sampling are available only
in the region $N  \leq 7$. The curve indicates a quadratic fit
to the results with the Coulomb gas representation given in 
Dean and Majumdar\cite{dean2008extreme}.}
  \label{fig:exceed0_GOE}
\end{center}
\end{figure}

Fig.~\ref{fig:exceed0_GOE} shows our numerical results for GOE.
We can treat matrices up to $N = 40$,
which is not treated by simple random sampling. 
Here we use the random walk scheme
$r_{ij}(x_{ij}^{new}|x^{old})=p_{ij}(x_{ij}^{new}-x_{ij}^{old})$
and the total number of matrix diagonalizations is
$3\times 10^9$ for $N=4 \sim 20$ and $4.5\times 10^9$
for $N=22 \sim 40$.
The results for $N \leq 7$ coincide with those  by
simple random sampling.
They are also consistent with the fit 
$-0.272N^2-0.493N+0.244$
of the numerical results  calculated 
in~\cite{dean2008extreme} 
with the Coulomb gas representation.
Hence, probabilities as tiny as $\sim 10^{-200}$ are estimated
by the proposed method and  agree well with the known results.

\begin{figure}
\begin{center}
\includegraphics{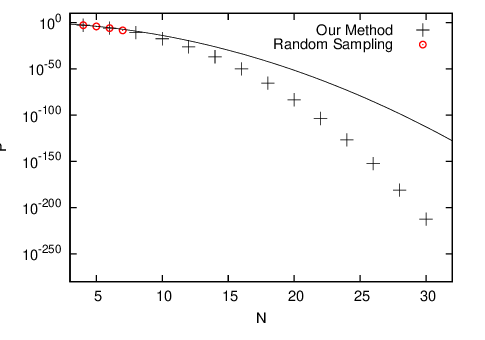}
\small
\caption{
Probability $P(\forall i, \lambda_i < 0)$  for an ensemble
of matrices whose components are uniformly distributed.
The horizontal axis corresponds to the size $N$ of the matrices.
The results of the proposed method (+) and the simple
random sampling method ($\odot$) are shown.
The results of simple random sampling are shown for
 $4 \leq N \leq 7$. 
The curve indicates the probability for the GOE with  the
same variance.
}
  \label{fig:exceed0_uniform}
\end{center}  
\end{figure}

Next, to show the flexibility of the proposed method, 
we calculate the probability 
$P(\forall i, \lambda_i < 0)$
for an ensemble of real symmetric matrices whose
entries are independently distributed with 
the uniform distribution $p_{ij}(x_{ij})$ defined by 
$$
p_{ij}(x_{ij})= \left\{
\begin{array}{ccc}
\frac{1}{2\sqrt{3}L} {}_,
 & \mbox{$|x_{ij}|<\sqrt{3}L$} & \mbox{and  $i=j$,} \\
\frac{1}{\sqrt{6}L} {}_, & \mbox{$|x_{ij}|<\sqrt{\frac{3}{2}}L$}
 & \mbox{and $i \neq j$,} \\
0 & \mbox{else.} &\\
\end{array}
\right.
$$
Hereafter, the value of the parameter $L$
is unity, which fits
the variances of the components 
to those of the GOE.
Results of the proposed method for this ensemble
are shown in Fig.~\ref{fig:exceed0_uniform}.
The proposal density 
$r_{ij}(x_{ij}^{new}|x^{old})=p_{ij}(x_{ij}^{new})$ is used.
Total number of matrix diagonalizations is
$4.5\times 10^9$ for each value of $N$; two third of which
are used to optimize the weight.
Fitting the results yields asymptotic behavior of
the probability,
$$
 P(\forall i, \lambda _i < 0) \sim \exp \left ( 
-aN^2-bN-c 
\right )
$$
for large $N$, where $a=0.679$, $-b=4.76$, and  $c=17.31$.
As shown in Fig.~\ref{fig:exceed0_uniform},
these probabilities significantly differ from that for
the GOE with the same variance.

\section{Sparse Random Matrices} \label{sec:sparse}

We also study ensembles of sparse random matrices.
Once the matrices become sparse, the Coulomb gas
approach is not applicable even in  Gaussian cases. 
The proposed approach allows 
us to calculate the probability $P(\forall i, \lambda_i < 0)$
in these cases. In this section, we use the proposal density 
$r_{ij}(x_{ij}^{new}|x^{old})=p_{ij}(x_{ij}^{new})$, but
the results are also checked by random walk schemes.

Various ways of defining  sparse random matrices are available.
Among them, we consider two types of  definitions in this study.
The first is as follows:
(1)~The matrix is symmetric.
(2)~All diagonal entries are $-1$. (3)~Nonzero off-diagonal entries
in the upper half of the matrix are mutually independent Gaussian 
variables  with zero mean and unit variance. (4)~Total number of nonzero entries is 
fixed at $\gamma N$, where $\gamma$ is the average number of nonzero 
entries per row. (5)~The positions of  nonzero off-diagonal entries 
in the upper half of the matrix  are randomly chosen.

\begin{figure}
\begin{center}
\includegraphics{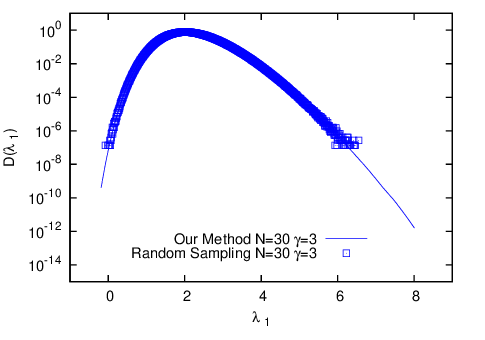}
\small
  \caption{
Density $D(\lambda_1)$ in a  case of  sparse random matrices.
The first definition is applied; results of the proposed method and 
the simple random sampling method are compared for 
$N=30$ and $\gamma=3$.
The symbol {\tiny $\Box$} appears only in the region where
simple random sampling gives nonzero results.  
}
  \label{fig:old_sparse}
\end{center}  
\end{figure}

The total number
of nonzero components should be preserved
with this definition. Hence,
the single component update in previous sections  is
replaced by a trial of exchanging zero and nonzero components 
with resampling of the nonzero component. 
Other parts of the algorithm remain essentially the same.
An example of the density $D(\lambda_1)$ computed
by this modified method is shown in Fig.~\ref{fig:old_sparse}.

\begin{figure}
\begin{center}
\includegraphics{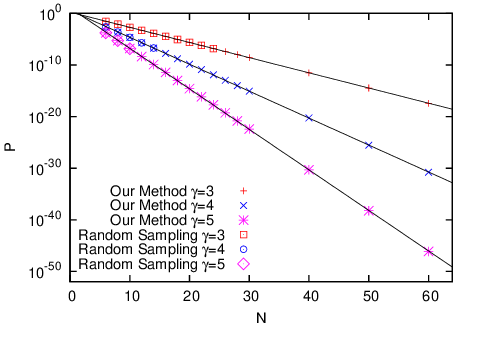}
\small
  \caption{
Probabilities $P(\forall i, \lambda_i < 0)$  for an ensemble
of  sparse random  matrices  estimated  by  the proposed method.  
The first definition is applied;  the results with $\gamma=3$, $4$, and $5$
versus size $N$ of the matrices are shown.
The lines show  linear fits of the data.
}
  \label{fig:exceed0_sparse}
\end{center}  
\end{figure}

The probability
$P(\forall i, \lambda _i < 0)$
that all eigenvalues are negative is also
successfully calculated by this algorithm  
for $\gamma=3,4$, and $5$, as shown in 
Fig.~\ref{fig:exceed0_sparse}.
These results indicate that for sparse random matrices,
the probability $P(\lambda _i < 0,\forall i)$ behaves as
$$
P(\forall i, \lambda _i < 0) \sim \exp \left ( -a _\gamma N  \right )
$$
for large $N$, where the estimated values of the
constants $a _\gamma$ are $a _3=0.68$, $a _4=1.20$, and $a_5 =1.81$
for $\gamma=3, 4$, and $5$, respectively.

In the case of sparse matrices, 
the log-probability $\log P(\forall i, \lambda _i < 0)$ 
is linear in $N$, which is apparently different from the behavior
proportional to  $N^2$ seen in the previous two examples.
However, if we plot 
the probability $P(\forall i, \lambda _i < 0)$ 
with the number $M$  of nonzero components 
instead of the size $N$, the dependence is linear 
in all examples.  Because $M \propto N$ in a sparse case and
$M \propto N^2$ in a dense case, the obtained results are 
naturally explained.

The definition of sparse random matrices 
most frequent in the literature~\cite{
rodgers1988density, *mirlin1991universality, *semerjian2002sparse}
differs from that given above.  Here, 
a second definition of sparse random matrices
is given by assigning the probability
$$
 p_{ij}(x_{ij})=\left (1-\frac{\gamma}{N} \right )\delta (x_{ij})+\frac{\gamma}{N}\pi (x_{ij})
$$
to all components $x_{ij}, \,\, \mbox{\small ($ i \geq j$)}$, where $\delta$ and $\pi$ 
denote Dirac's delta function and a Gaussian density
with zero mean and unit variance, respectively; each component is 
assumed to be an independent sample from this distribution.

In this case, all components are mutually independent and the modification
for  keeping  the number of nonzero components is not necessary.
However, since the  diagonal elements can vanish, 
singular behavior of the density $D(\lambda_1)$  of states
appears at $\lambda_1=0$, which affects the efficiency of the proposed
method.  

Fortunately, when we are interested in $P(\forall i, \lambda _i < 0)$,
this difficulty is easily treated; we use the fact that the  condition 
that all diagonal elements are negative \mbox{$\forall i,  x_{ii}<0$} 
is a necessary condition for \mbox{$\forall i,  \lambda _i < 0$}.
By using this condition, the following
two-stage method is introduced. 
First, we calculate the conditional 
probability \mbox{$P(\forall i, \lambda _i < 0 \, | \, \forall i,  x_{ii}<0)$}.
This conditional probability can be calculated  with a multicanonical algorithm, 
in which we reject any state $\exists i,  x_{ii} \geq 0$.
The second step is to calculate the probability
\mbox{$P( \forall i,  x_{ii}<0)$}. Elementary calculation shows
that 
\begin{equation} \label{eq:relation}
P(\forall i,  x_{ii}<0) = \left(\frac{\mathstrut 1}{ \mathstrut2}\right)^N \times \left 
(\frac{\mathstrut \gamma}{\mathstrut N} \right )^N.
\end{equation}
Then, the probability $P(\forall i, \lambda _i < 0)$ is given by
the product
\mbox{$P(\forall i, \lambda _i < 0 \, | \, \forall i,  x_{ii}<0)
\times P(\forall i,  x_{ii}<0)$}.

\begin{figure}
\begin{center}
\includegraphics{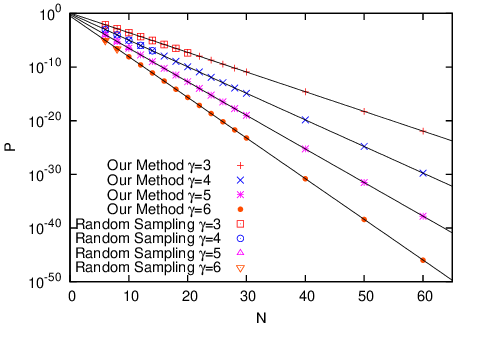}
\small
  \caption{
Probabilities $P(\forall i, \lambda _i < 0 \, | \, \forall i,  x_{ii}<0)$ 
 for an ensemble of   sparse random matrices  
estimated  by  the proposed method. 
The second definition is applied;  the results with $\gamma=3$, $4$,
$5$, and $6$ versus size $N$ of the matrices are shown. 
The lines show  linear fits to the data.
The results of simple random sampling are also shown.
}
  \label{fig:exceed00new_sparse}
\end{center}  
\end{figure}

\begin{figure}[thb]
\begin{center}
\includegraphics{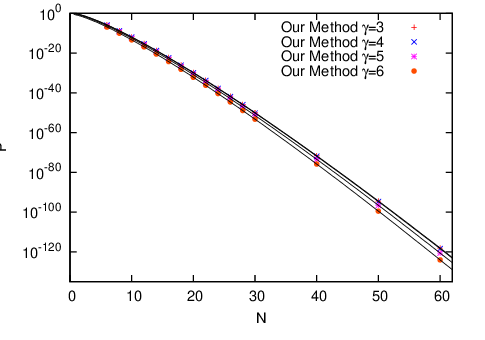}
\small
  \caption{
Probabilities $P(\forall i, \lambda_i < 0)$  for an ensemble
of   sparse random matrices  estimated  by  the proposed method. 
The second definition is applied;  the results with $\gamma=3$, $4$,
$5$, and $6$ versus size $N$ of the matrices are shown. 
The lines show fits using (\ref{eq:spfit}).}
  \label{fig:exceed0new_sparse}
\end{center}  
\end{figure}

Fig.~\ref{fig:exceed00new_sparse} shows examples of 
the probability
\mbox{$P(\forall i, \lambda _i < 0 \, | \, \forall i,  x_{ii}<0)$}
calculated in the first step; 
it  is  linear in $N$ in the semi-log scale, as expected. 
The probability \mbox{$P(\forall i, \lambda _i < 0)$} obtained from it
is shown  in Fig.~\ref{fig:exceed0new_sparse}.  
In this case, \mbox{$\log P( \forall i, \lambda _i < 0)$} is 
no longer linear in $N$ because of  an $O(N\log N)$ term  
arising from (\ref{eq:relation}). They are fitted as 
\begin{equation}\label{eq:spfit}
P(\forall i, \lambda _i < 0) \sim 
 \left (\frac{\gamma}{2N} \right )^N \exp(-a_{\gamma} N)_,
\end{equation}
where $a_3=0.845$, $a_4=1.14$, $a_5=1.44$, and $a_6=1.75$
 for $\gamma=3, 4, 5$, and $6$, respectively. \\

\section{Concluding Remarks} \label{sec:con}

A  method based on multicanonical Monte Carlo 
is proposed and applied  to the estimation of 
large deviations in the largest eigenvalue of random matrices.
The method is successfully tested 
with the Gaussian orthogonal ensemble (GOE), 
an ensemble of matrices whose components are 
uniformly distributed in an interval, and
an ensemble of sparse random matrices.
The probabilities that all
eigenvalues of a matrix are negative are successfully
estimated in cases where simple random sampling 
is largely ineffective; the smallest values of the obtained
probabilities  are  $\sim 10^{-200}$.

The method can be applied to any ensemble
of matrices. Moreover,  it enables sampling of rare
events defined by any statistics.
Hence, it will be interesting to apply the method to
large deviations in other quantities, such as
statistics involving eigenvectors or 
spacing of eigenvalues.

\begin{acknowledgments}
We thank Prof.~M.~Kikuchi for his support
and encouragement.
This work is supported by Grants-In-Aid for 
Scientific Research (KAKENHI, No.17540348 and No.18079004) from MEXT of
 Japan. This work is also supported in part by Global COE Program (Core
 Research and Engineering of Advanced Materials-Interdisciplinary
 Education Center for Materials Science), MEXT, Japan. All simulations
 were performed on a PC cluster at Cybermedia center, Osaka university.
\end{acknowledgments}

\appendix*
\section{Wang-Landau algorithm}
Here we give a brief account of the algorithm
used in the paper. 
After the weight is tuned  by the procedure 
described below, the final long run using the weight gives
the estimate of the desired density, as explained in the main text. 
Detailed studies on the Wang-Landau algorithm
and other methods for realizing multicanonical weights
are found in references~\cite{lee2006convergence, malakis2006monte, 
berg1996multicanonical}.\\

\noindent
{\bf constants and arrays}\\ 
$\lambda_{\max}, \lambda_{\min}$ are the upper/lower bounds of $\lambda_1$.\\
$K$ is an integer that defines the number of bins.\\
$\Ww$ is a real array whose indices are in $1 \ldots K$.\\
$\Hh$ is an integer array whose indices are in $1 \ldots K$.\\
$\varphi$ is a function that maps: $\lambda_1 \rightarrow$ [indices of $\Ww$ and $\Hh$].\\
$f_0$ is an initial modification factor: $f_0=e=2.718$.\\
$b$ is a flatness constant: $b=0.92$.\\
$N_{out}$ is the number of iteration (outer loop) \\ : typically $10^3\sim 10^5$.\\
$N_{in}$ is the number of iteration (inner loop) \\ : typically $50000$.\\  

\noindent
{\bf initialization} \\
$f \leftarrow f_0$.\\
$n_c \leftarrow$ 1.\\
$\Ww(k) \leftarrow 1; \, k=1 \ldots K$.\\
$\Hh(k) \leftarrow 0; \, k=1 \ldots K$.\\
$\{x_{ij}\} \leftarrow$ arbitrary values. \\
$\lambda_1^c \leftarrow \lambda_1(\{x_{ij})\})$. \\

\noindent
{\bf outer loop} \\
Repeat the following steps $N_{out}$ times.\\

\noindent
\hspace*{0.4cm} {\bf inner loop}  \\
\hspace*{0.4cm} Repeat the following steps $N_{in}$ times.
\begin{itemize}
\item
Select an index $(i,j), \mbox{\small$i \geq j$}$ randomly.
\item
Generate  $x_{ij}^{new}$ as a sample from $r_{ij}(x_{ij}^{new}|\{x_{ij}\})$.
\item
$x^{new}_{lm} \leftarrow x_{lm}; \  \mbox{\small $(l,m) \neq (i,j)$}$.
\item 
$\lambda_1^{new} \leftarrow \lambda_1(\{x^{new}_{ij}\})$.
\item 
If  $\lambda_1^{new} \notin [\lambda_{\min},\lambda_{\max} ]$ jump to~\verb+*+.

\item
Calculate the ratio $R$ using
$$
R \leftarrow 
\frac{p_{ij}(x_{ij}^{new})}{p_{ij}(x_{ij})} 
\frac{r_{ij}(x_{ij}|\{x_{ij}^{new}\})}{r_{ij}(x_{ij}^{new}|\{x_{ij}\})}
\frac{\Ww(\varphi((\lambda_1^{new}))}{\Ww(\varphi(\lambda_1^c))}.
$$
\item
Generate a uniform random  number $u \in (0,1]$.
\item
If $u<R$, $x_{ij} \leftarrow x_{ij}^{new}$,\, $\lambda_1^c\leftarrow \lambda_1^{new}$.
\item
$\Ww(\varphi(\lambda_1^c)) \leftarrow 
\Ww(\varphi(\lambda_1^c)) \times 1/f $.  \ \verb+(*)+
\item
$\Hh(\varphi(\lambda_1^c)) \leftarrow 
\Hh(\varphi(\lambda_1^c))+1 $.
\end{itemize}
\vspace{-0.3cm}
\noindent
\hspace*{0.4cm} {\bf  end of inner loop;}\\

\noindent
\ $\bar{\Hh} \leftarrow \sum_{i=1}^K \Hh(k) /K$. \\

\noindent
\hspace*{1ex}If \ ($\bar{\Hh}*b \leq \Hh(k)$ ; $k=1\ldots K$) \ then\\
\ \ \ \ $f \leftarrow \sqrt{f}$. \\
\ \ \ \ $\Hh(k) \leftarrow 0; \, k=1 \ldots K$.\\
\ \ \ \ $n_c  \leftarrow n_c+1$.\\
\hspace*{1ex}end if \\

\noindent
{\bf end of outer loop;} \\

A few remarks on the algorithm are in order:
\begin{itemize}
\item
Arrays representing the weight 
$w$ and histogram $h$ are denoted by $W$ and $H$, respectively.

\item
Maximum eigenvalue $\lambda_1(x)$ of the matrix $x$ is
calculated by the Householder algorithm at each Metropolis-Hastings
step, which is the most time consuming part of the algorithm.
\item
In the current implementation, we fix the number 
$N_{out}$ of iteration of outer loop; if the value of $f$ is
not sufficiently close to the unity at the end of 
computation (i.e., $n_c \lesssim 18$),
we repeat it from the beginning with an increased $N_{out}$.
This is enough for our purpose of testing the algorithm,
although sophisticated stopping criteria could save 
the computational time.
\item
When the support of the target density 
$D(\lambda)$ is unbounded, the choice
of the bounds $\lambda_{\max}$ and $\lambda_{\min}$
significantly affects the computational time; if we want to calculate
extreme tails, $N_{out}$ and $N_{in}$
defined above should be large to ensure
$f \simeq 1$ at the end of the computation.
The interval $[\lambda_{\max}, \lambda_{\min}]$
should contain $\lambda_1(x)$ for 
the initial value of $x$; usually it is easy to realize.
\end{itemize}

%

\end{document}